\documentclass[apex, twocolumn]{jjap3_arXiv}
\usepackage{txfonts}

\usepackage[format=hang,font=footnotesize]{caption}

\makeatletter
\def\ext@figure{}
\makeatother
\title{Strain-induced large spin splitting and persistent spin helix at LaAlO$_3$/SrTiO$_3$ interface}
\author{Naoya Yamaguchi$^{1}$\thanks{E-mail: n-yamaguchi@cphys.s.kanazawa-u.ac.jp} and Fumiyuki Ishii$^{2}$\thanks{E-mail: ishii@cphys.s.kanazawa-u.ac.jp}}
\inst{$^{1}$Division of Mathematical and Physical Sciences, Graduate School of Natural Science and Technology, Kanazawa University, Kanazawa, 920-1192, Japan.\\
$^{2}$Faculty of Mathematics and Physics, Institute of Science and Engineering, Kanazawa University, Kanazawa, 920-1192, Japan.}
\abst{We investigated the effect of the tensile strain on the spin splitting at the n-type interface in LaAlO$_3$/SrTiO$_3$ in terms of the spin-orbit coupling coefficient $\alpha$ and spin texture in the momentum space using first-principles calculations.
We found that the $\alpha$ could be controlled by the tensile strain and be enhanced up to 5 times for the tensile strain of 7\%, and the effect of the tensile strain leads to a persistent spin helix, which has a long spin lifetime.
These results support that the strain effect on LaAlO$_3$/SrTiO$_3$ is important for various applications such as spinFET and spin-to-charge conversion.}

\begin{document}
\maketitle
The LaAlO\textsubscript{3}/SrTiO\textsubscript{3} interface is one of
the most important heterostructures, which has an n-type interface with
high mobility two-dimensional electronic gas (2DEG)
\cite{Ohtomo_A_2004, Thiel_Tunable_2006}. Recently, Rashba effect in
LaAlO\textsubscript{3}/SrTiO\textsubscript{3} has attracted much
attention due to potential applications in spintronics.
\cite{Caviglia_Tunable_2010}. For example, spin-to-charge conversion
phenomenon induced by the inverse Rashba--Edelstein effect (IREE) for
LaAlO\textsubscript{3}/SrTiO\textsubscript{3} was reported
\cite{Lesne_Highly_2016}. IREE can be used for an alternative mechanism
of detection of spin current utilizing inverse spin Hall effect
\cite{Saitoh_Conversion_2006}, and it is suggested that the strength of
the IREE was proportional to the Rashba coefficient \(\alpha_R\)
\cite{Snchez_Spin_2013}. Therefore it is quite important to control the
Rashba effect at the LaAlO\textsubscript{3}/SrTiO\textsubscript{3}
interface.

The applied electric field can control the magnitude of \(\alpha_R\).
The \(\alpha_R\) in LaAlO\textsubscript{3}/SrTiO\textsubscript{3}
induced by built-in electric field was experimentally evaluated as 18
meV\(\cdot\AA\) and it was increased up to 49 meV\(\cdot\AA\) by applying
electric field \cite{Caviglia_Tunable_2010}. In epitaxial oxide thin
films and superlattices, strain induced by substrates, is useful to
control electronic structures and physical properties, and so important
that its effects are extensively studied for various materials
\cite{Diguez_First_2005}. Some works about the strain for
LaAlO\textsubscript{3}/SrTiO\textsubscript{3} actually were reported
\cite{Bark_Tailoring_2011, Doennig_Control_2015}. Bulk
SrTiO\textsubscript{3} shows strain-induced ferroelectric phase
transition \cite{Haeni_Room_2004}, and the electric polarization is
induced by the tensile strain along the {[}110{]}-direction. Since the
2DEG at the n-type interface in
LaAlO\textsubscript{3}/SrTiO\textsubscript{3} may be influenced by the
strain-induced electric polarization, the spin splitting, especially
spin-orbit coupling coefficient \(\alpha\), can be controlled by the
strain. In fact, a first-principles calculation of strain control of the
Rashba spin splitting in ZnO was reported \cite{Absor_Tunable_2014}.

In this work, we focus the spin splitting at the n-type interface in
LaAlO\textsubscript{3}/SrTiO\textsubscript{3}, and investigate the
effect of tensile strain on it in terms of the spin-orbit coupling
coefficient \(\alpha\) and spin texture using first-principles
calculations. We found that the \(\alpha\) could be controlled by the
tensile strain and be enhanced up to 5 times for the tensile strain of
7\%, and the effect of the tensile strain leads to a persistent spin helix,
which has a long spin lifetime. These properties are expected to be
important to various spintronic applications.

Rashba effect \cite{Bychkov_Properties_1984} arises from spin-orbit
interaction for the 2DEG at the surfaces and interfaces with the spatial
inversion symmetry breaking, and the Hamiltonian describing Rashba
effect in the 2DEG can be expressed by
\(H=-\hbar^2\nabla_{\parallel}^2/(2m^*)+H_R\), where \(\hbar\) is the
Planck constant, and \(m^*\) is the effective mass of electrons. The
Rashba Hamiltonian:
\(H_R=\alpha_R(\hat{e}_z\times\vec{k}_{\parallel})\cdot\vec{\sigma}=\alpha_R(k_y\sigma_x-k_x\sigma_y)\),
where \(\alpha_R\) denotes the Rashba coefficient, \(\hat{e}_{z}\) the
unit vector along \(z\)-axis, \(\vec{k}_{\parallel}=(k_x, k_y, 0)\) the
wave vector, and \(\vec{\sigma}=(\sigma_x, \sigma_y, \sigma_z)\) the
Pauli matrices vector, respectively. We can obtain the energy dispersion
relation:
\(E_\pm(\vec{k}_{\parallel})=\hbar^2k_{\parallel}^2/(2m^*)\pm\alpha_R k_{\parallel}.\)
The \(\alpha_R\) satisfies \(\alpha_R=2E_R/k_R\), where \(E_R\)
(\(=m^*\alpha_R^2/(2\hbar^2)\)) is the Rashba energy, and \(k_R\)
(\(=m^*\alpha_R/\hbar^2\)) the Rashba momentum offset.

For the interfacial system with the tensile strain in
LaAlO\textsubscript{3}/SrTiO\textsubscript{3}, due to the strain-induced
electric polarization along the {[}110{]}-direction originating from
bulk SrTiO\textsubscript{3}, there is expected to be only one mirror
plane along the \((\bar110)\) plane. Indeed, we confirm that atomic
displacement is induced along the \((\bar110)\) plane in atomic
structures optimized by the first-principles calculation. Thus, the
spin-orbit Hamiltonian can be expressed as
\(H_{SO}=\alpha_{xy}^{\parallel}k_x\sigma_y+\alpha_{yx}^{\parallel}k_y\sigma_x+\alpha_{[1\bar10]z}^\perp\left(\frac{k_x-k_y}{\sqrt2}\right)\sigma_z\)
in the same way as the analysis in the previous work of a ZnO
\((10\bar10)\) surface with a mirror plane along the (100) plane
\cite{Absor_Persistent_2015}. While the \(\alpha_{xy}^\parallel\) and
\(\alpha_{yx}^\parallel\) are corresponding to the Rashba effect, the
\(\alpha_{[1\bar10]z}^\perp\) originates from the strain-induced
electric polarization. As the tensile strain increases,
\(H_{[1\bar10]z}^\perp=\alpha_{[1\bar10]z}^\perp\left(\frac{k_x-k_y}{\sqrt2}\right)\sigma_z\)
is expected to be more dominant. The spin texture for
\(H_{[1\bar10]z}^\perp\) has the spin splitting along the
\([1\bar10]\)-direction, and the \(\alpha_{[1\bar10]z}^\perp\) can be
evaluated in the same way as \(\alpha_R\) for the Rashba system:
\(\alpha_{[1\bar10]z}^\perp=2E_{[1\bar10]z}^\perp/k_{[1\bar10]z}^\perp\).
We take the \(\alpha_R\) and \(\alpha_{[1\bar10]z}^\perp\) as
\(\alpha_{[1\bar10]}\) hereafter.

\begin{figure}
\begin{minipage}[t]{0.18\textwidth}
\centering\strut
\begin{enumerate}
\def\labelenumi{(\alph{enumi})}
\setcounter{enumi}{0}
\item
\end{enumerate}
\includegraphics[width=\columnwidth]{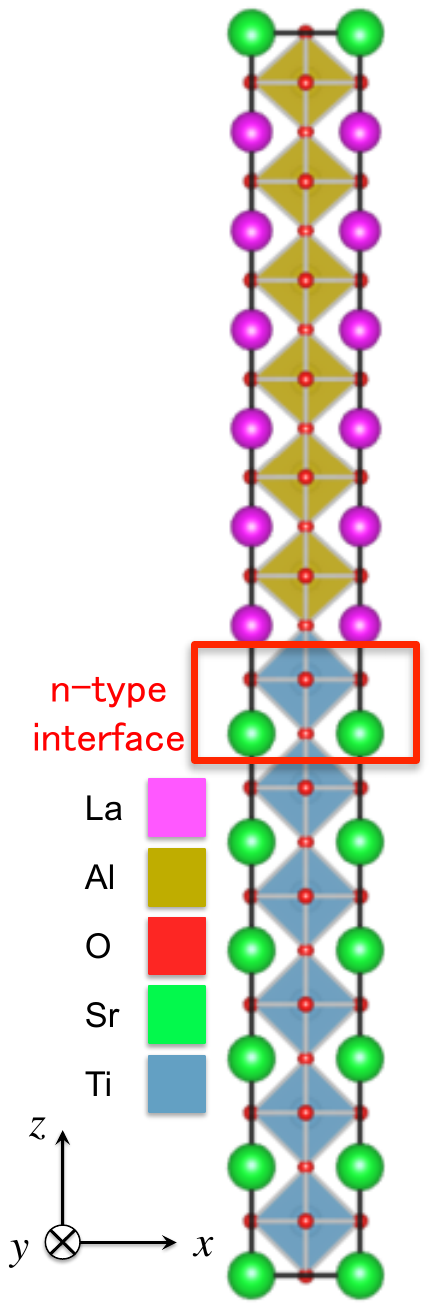}
\strut
\vspace{-4ex}
\end{minipage}
\begin{minipage}[t]{0.28\textwidth}
\centering\strut
\begin{enumerate}
\def\labelenumi{(\alph{enumi})}
\setcounter{enumi}{1}
\item
\end{enumerate}
\includegraphics[width=0.5\columnwidth]{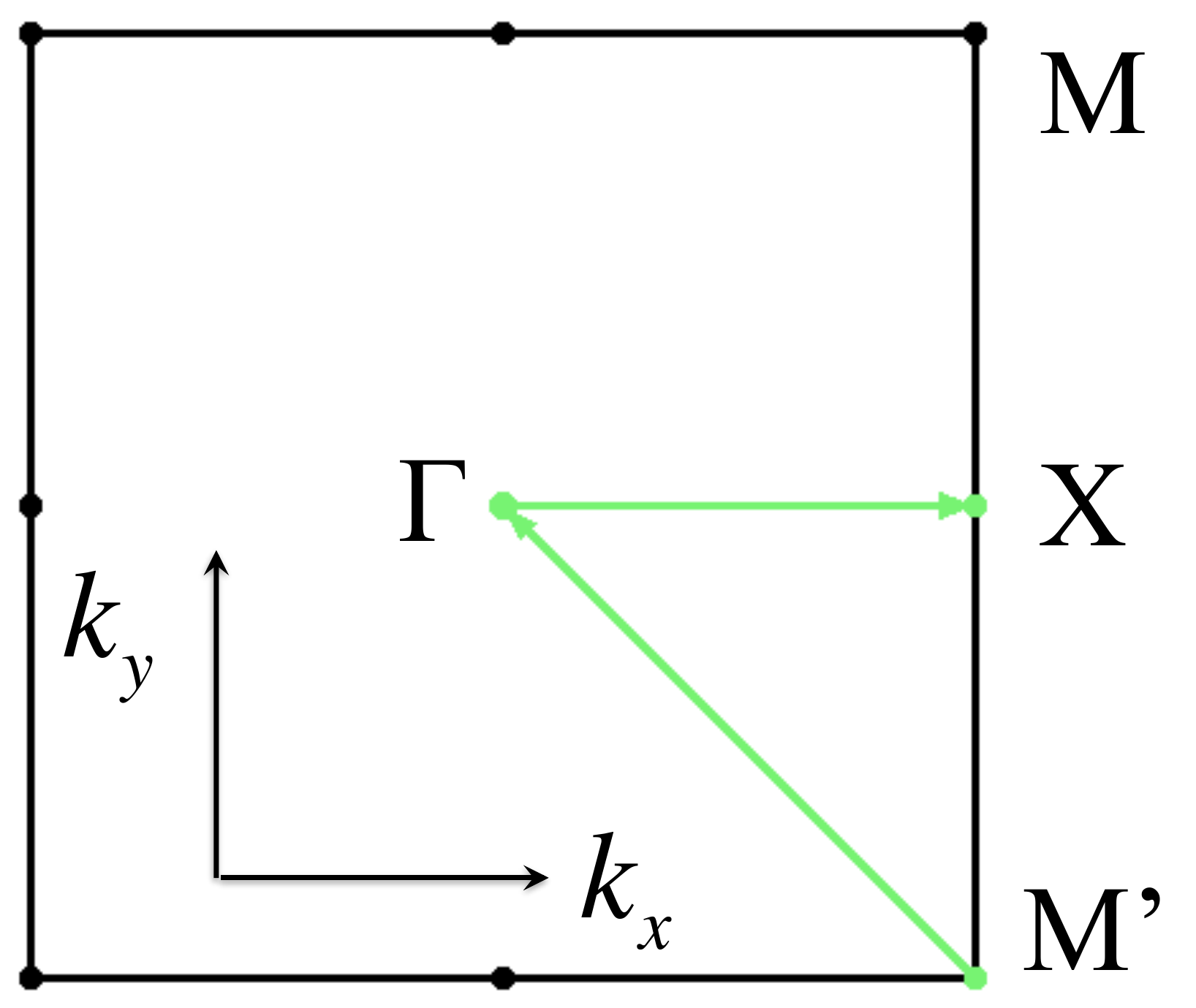}
\begin{enumerate}
\def\labelenumi{(\alph{enumi})}
\setcounter{enumi}{2}
\item
\end{enumerate}
\includegraphics[width=\columnwidth]{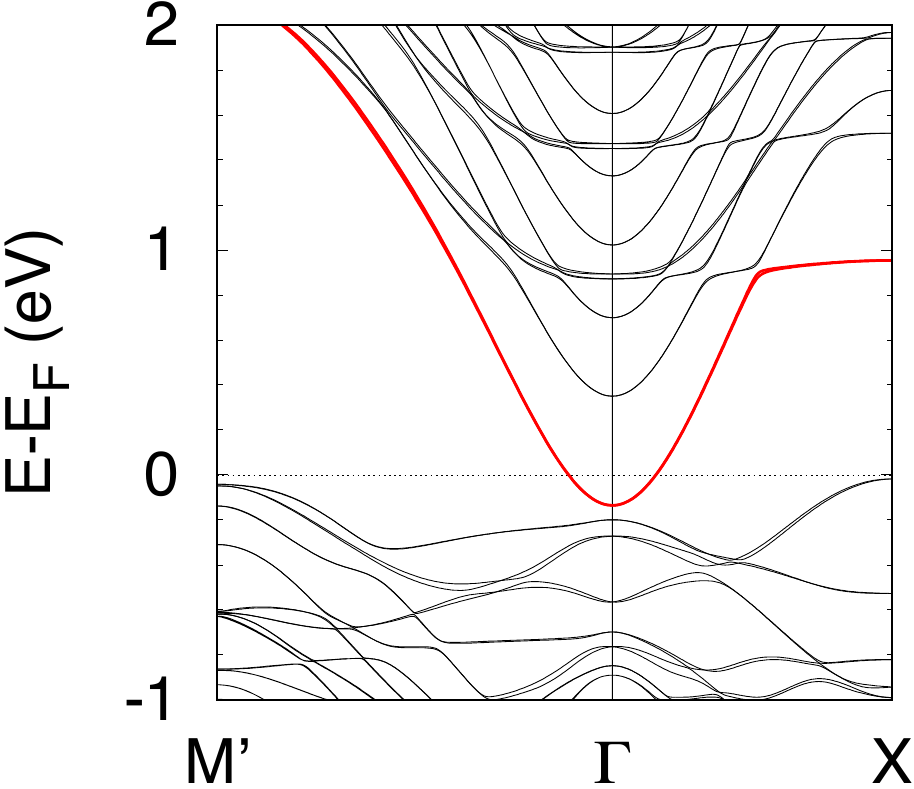}
\begin{enumerate}
\def\labelenumi{(\alph{enumi})}
\setcounter{enumi}{3}
\item
\end{enumerate}
\includegraphics[width=\columnwidth]{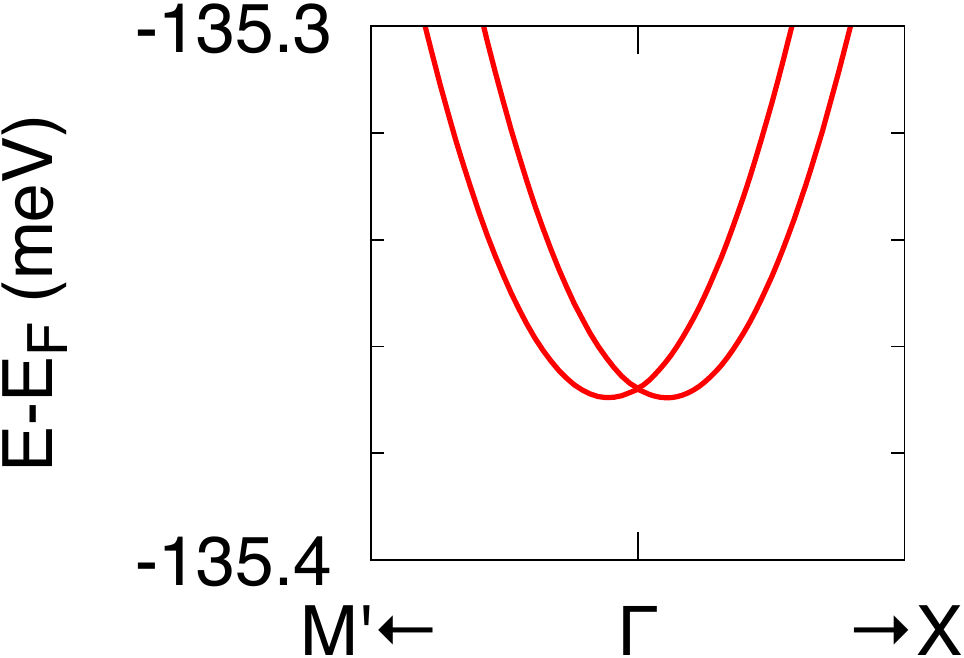}
\strut
\vspace{-4ex}
\end{minipage}
\caption{(a) Computational model. (b) Schematic of the first Brillouin zone. (c) Band structure for unstrained LaAlO$_3$/SrTiO$_3$. The red curves show the bands with the Rashba spin splitting. (d) Enlarged view for the interfacial bands around $\Gamma$-point.}\tabularnewline
\vspace{-6ex}
\end{figure}

We perform density functional calculations using the computational model
of the superlattice
(LaAlO\textsubscript{3})\textsubscript{6}/(SrTiO\textsubscript{3})\textsubscript{6}
shown in Fig. 1(a), where the n-type interface indicates the one between
the TiO\textsubscript{2} and LaO layers. It was suggested that the
number of the formula unit of LaAlO\textsubscript{3} or
SrTiO\textsubscript{3} was required to be 6 to describe the metallic
ground state with atomic relaxation
\cite{Ishibashi_Analysis_2008, Park_Charge_2006, Pentcheva_Avoiding_2009, Nishida_First_2014}.
We use the experimental lattice constants: \(a_0^{LAO}\)=3.788 \(\AA\)
\cite{savchenko1985phase} for LaAlO\textsubscript{3};
\(a_0^{STO}\)=3.905 \(\AA\) \cite{Inaguma_Quantum_1992} for
SrTiO\textsubscript{3}, and the supercell length \(c\) is determined
with the cell volume conserved. We assumed the biaxial misfit strain
\((a−-a_0^{STO})/a\), where \(a\) denotes the in-plane lattice constant
(\(xy\)-plane), and the positive strain is tensile. Our calculations are
performed within the general gradient approximation
\cite{Perdew_Generalized_1996} by OpenMX code \cite{OpenMX}, with the
fully relativistic total angular momentum dependent pseudopotentials
taking spin-orbit interaction (SOI) into account
\cite{Theurich_Self_2001}. We adopt norm-conserving pseudopotentials
with an energy cutoff of 300 Ry for charge density including the 5\(s\),
5\(p\), 5\(d\) and 6\(s\)-states as valence states for La; 3\(s\) and
3\(p\) for Al; 2\(s\) and 2\(p\) for O; 4\(s\), 4\(p\) and 5\(s\) for
Sr; 3\(s\), 3\(p\), 3\(d\) and 4\(s\) for Ti. We use \(8\times8\times1\)
regular k-point mesh. The numerical pseudo atomic orbitals are used as
follows: the numbers of the \(s\)-, \(p\)-, \(d\)- and \(f\)-character
orbitals are 3, 3, 1 and 1, respectively, for La; 3, 3, 1 and 0 for Al
and O; 3, 2, 1 and 0 for Sr; 3, 3, 2 and 0 for Ti. The cutoff radii
of La, Al, O, Sr and Ti are 6.0, 7.0, 5.0, 10.0 and 7.0, respectively,
in units of Bohr.

The calculated band structure for the unstrained system is shown in Fig.
1(c), and Fig. 1(d) is the enlarged view of the interfacial bands that
we focus on. Analyzing the partial density of states (PDOS), we confirm
that the interfacial band, that is, the conduction band edge (CBE) in
the n-type interface mainly consist of the 3\(d_{xy}\) orbital as
reported in the previous study \cite{Du_The_2015}. For the tensile
strain, the CBE is below the Fermi level so that the interface is
metallic. This implies that the 2DEG remains for the tensile strain.

\begin{figure}
\begin{minipage}[t]{0.24\textwidth}
\centering\strut
\begin{enumerate}
\def\labelenumi{(\alph{enumi})}
\setcounter{enumi}{0}
\item
\end{enumerate}
\includegraphics[width=\columnwidth]{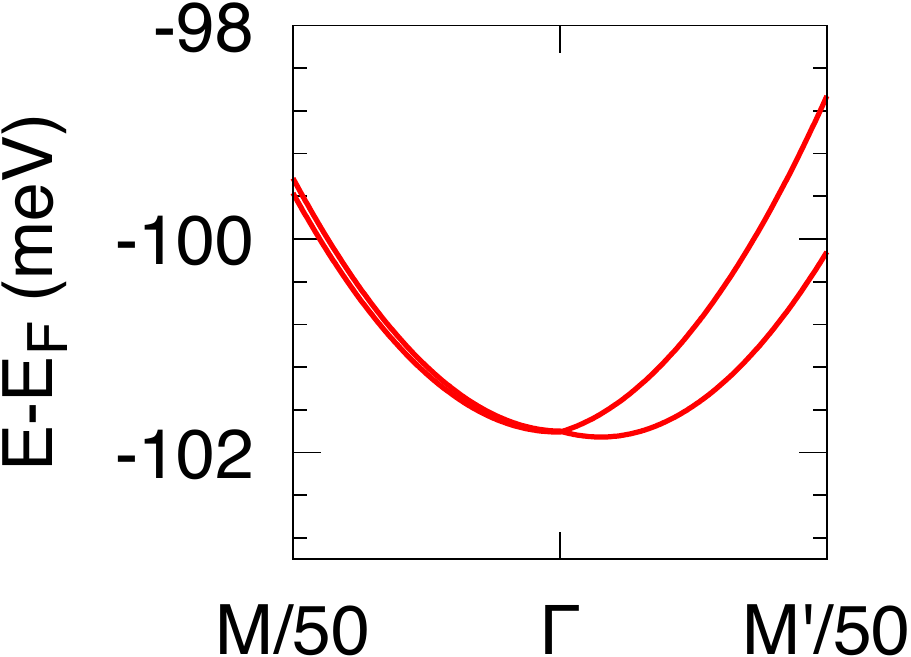}
\strut
\vspace{-5ex}
\end{minipage}
\begin{minipage}[t]{0.24\textwidth}
\centering\strut
\begin{enumerate}
\def\labelenumi{(\alph{enumi})}
\setcounter{enumi}{1}
\item
\end{enumerate}
\includegraphics[width=\columnwidth]{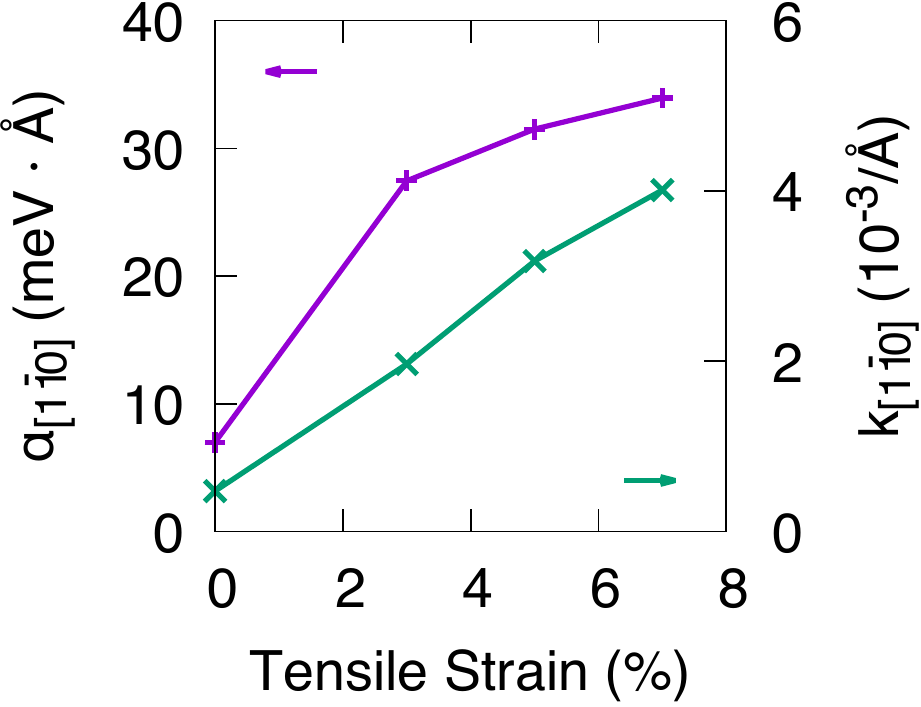}
\strut
\vspace{-5ex}
\end{minipage}
\begin{minipage}[t]{0.24\textwidth}
\centering\strut
\begin{enumerate}
\def\labelenumi{(\alph{enumi})}
\setcounter{enumi}{2}
\item
\end{enumerate}
\includegraphics[width=\columnwidth]{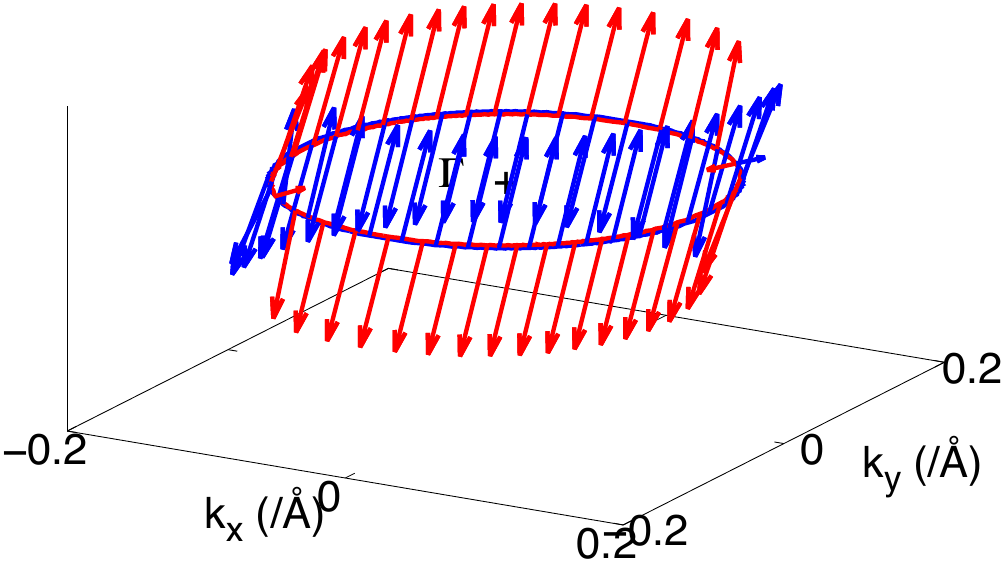}
\strut
\vspace{-4ex}
\end{minipage}
\begin{minipage}[t]{0.24\textwidth}
\centering\strut
\begin{enumerate}
\def\labelenumi{(\alph{enumi})}
\setcounter{enumi}{3}
\item
\end{enumerate}
\includegraphics[width=\columnwidth]{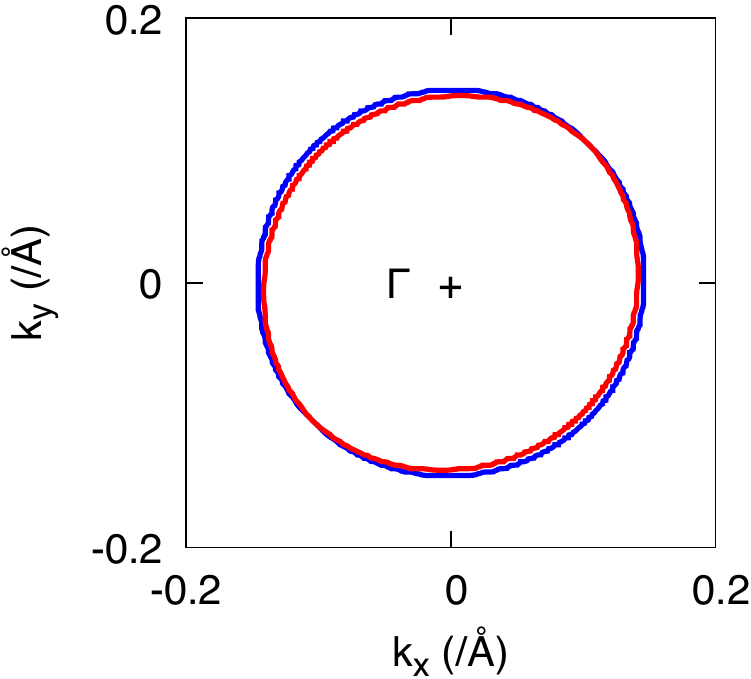}
\strut
\vspace{-4ex}
\end{minipage}
\caption{(a) Enlarged view for the interfacial bands about the spin splitting at the tensile strain of 5\%. (b) Strain dependence for the spin-orbit coefficient $\alpha_{[1\bar10]}$ and the momentum offset $k_{[1\bar10]}$. (c) Spin textures at the tensile strain of 5\%. (d) Fermi arcs at the tensile strain of 5\%.}\tabularnewline
\vspace{-6ex}
\end{figure}

We calculate the spin-orbit coupling coefficient \(\alpha_{[1\bar10]}\)
along the \([1\bar10]\)-direction (\(\Gamma\to\)M') at the n-type
interface for the spin splitting under the tensile strain inducing the
2DEG. With the tensile strain, we find larger spin splitting compared to
the unstrained system as shown in Fig. 2(a). Figure 2(b) shows the
strain dependence for the \(\alpha_{[1\bar10]}\) and the momentum offset
\(k_{[1\bar10]}\) for the interfacial bands around \(\Gamma\)-point.
Without strain, our calculated \(\alpha_{[1\bar10]}\) (7.49
meV\(\cdot\AA\)) is of the same order as the prior works (18 meV\(\cdot\AA\)
(expt.) \cite{Caviglia_Tunable_2010}; 12.6 meV\(\cdot\AA\) (theor.)
\cite{Nishida_First_2014}). As the tensile strain increases, the
magnitude of the \(\alpha_{[1\bar10]}\) increases. The strain-induced
electric polarization from bulk SrTiO\textsubscript{3} may enhance the
polarity along the {[}110{]}-direction
(\(\Gamma\to\)M), which makes \(\alpha_{[1\bar10]}\) larger. The
\(\alpha_{[1\bar10]}\), that is, spin splitting can be controlled by the
tensile strain to get the about 5 times larger \(\alpha_{[1\bar10]}\)
for the tensile strain of 7\%. The spin textures in the strained system
may be effective for IREE since it shows the pairs of the opposite spin
components in each Fermi arc: inner and outer ones (See Fig. 2(c), (d)),
while it is not Rashba spin splitting. The magnitude of the
\(\alpha_{[1\bar10]}\) as well as \(\alpha_R\) can be considered to be
directly connected to the efficiency of spin-to-charge conversion
\(\lambda_{IREE}\) due to the relation
\(\lambda_{IREE}=\alpha_R\tau_s/\hbar\) \cite{Snchez_Spin_2013}, where
\(\tau_s\) is the spin relaxation time. The effect of the tensile strain
may also enhance the \(\lambda_{IREE}\) in
LaAlO\textsubscript{3}/SrTiO\textsubscript{3} up to 5 times so that the
larger conversion may be realized than the previous work
\cite{Lesne_Highly_2016}.

\begin{figure}
\begin{minipage}[t]{0.24\textwidth}
\centering\strut
\begin{enumerate}
\def\labelenumi{(\alph{enumi})}
\setcounter{enumi}{0}
\item
\end{enumerate}
\includegraphics[width=\columnwidth]{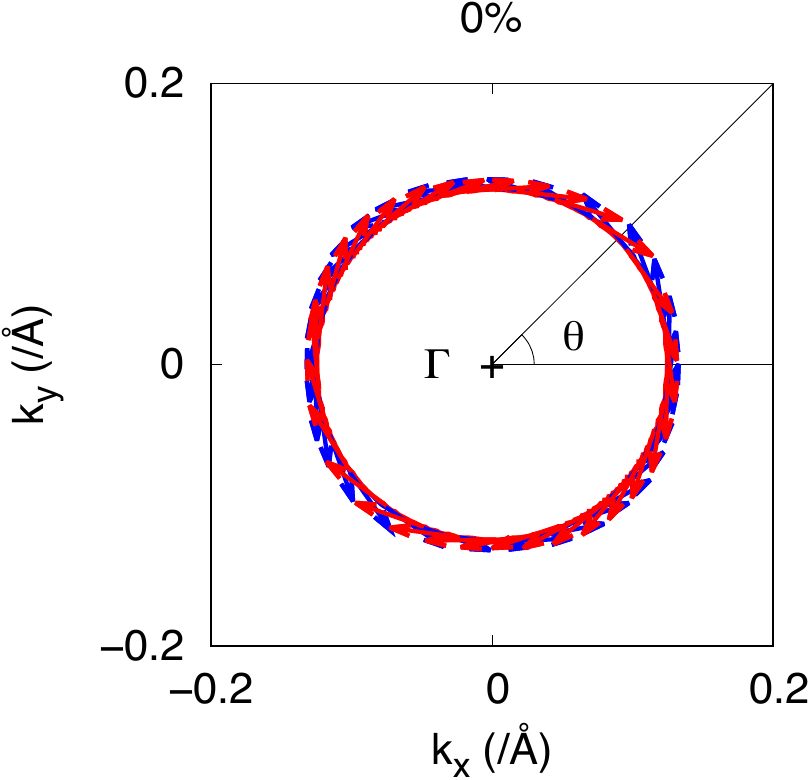}
\strut
\vspace{-5ex}
\end{minipage}
\begin{minipage}[t]{0.24\textwidth}
\centering\strut
\begin{enumerate}
\def\labelenumi{(\alph{enumi})}
\setcounter{enumi}{1}
\item
\end{enumerate}
\includegraphics[width=\columnwidth]{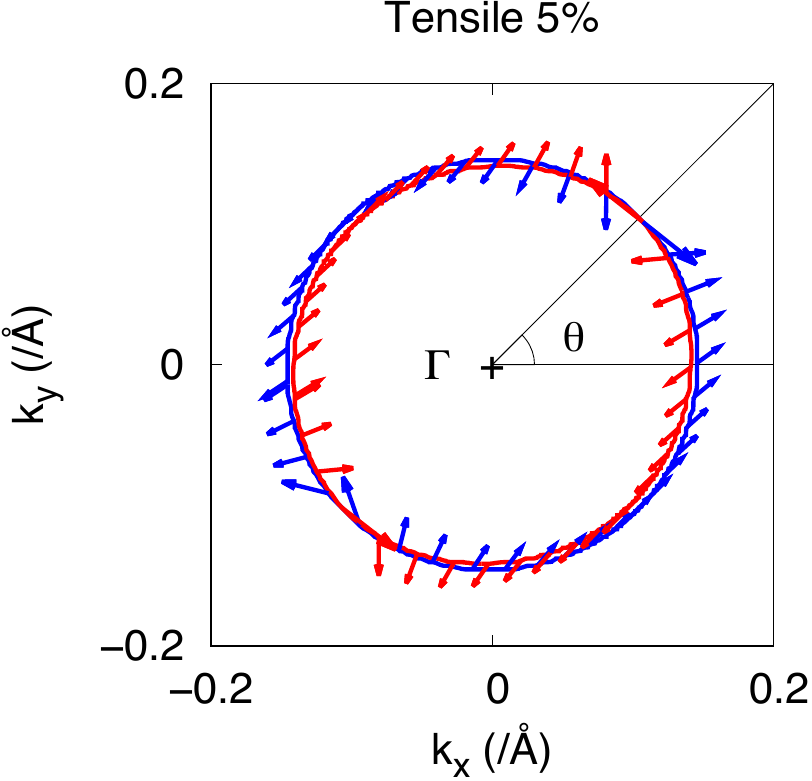}
\strut
\vspace{-5ex}
\end{minipage}
\centering
\begin{minipage}[t]{0.48\textwidth}
\centering\strut
\begin{enumerate}
\def\labelenumi{(\alph{enumi})}
\setcounter{enumi}{2}
\item
\end{enumerate}
\includegraphics[width=\columnwidth]{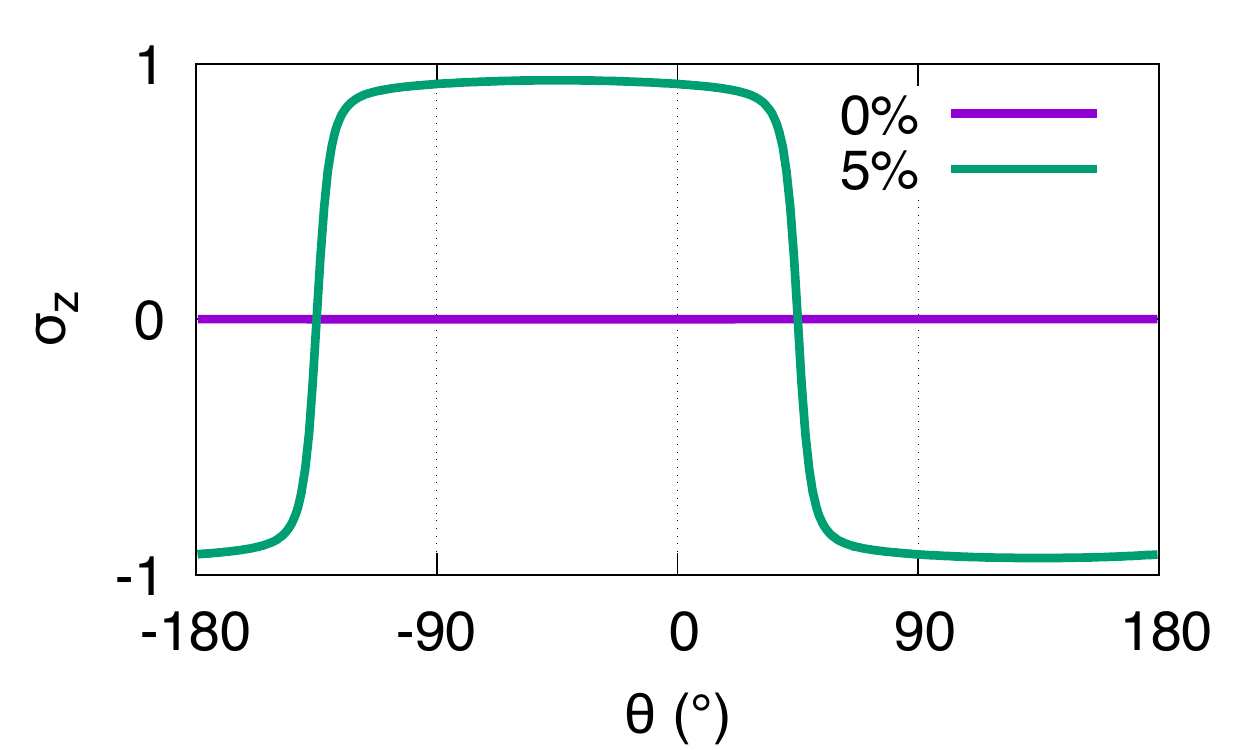}
\strut
\vspace{-4ex}
\end{minipage}
\caption{(a), (b): In-plane spin textures for the Rashba spin splitting at the Fermi level for each strain: (a) 0\%; (b) Tensile 5\%. The blue (red) arrow shows the in-plane components of the Pauli matrices vector for the outer (inner) Fermi arc. (c): Angular dependence for the out-of-plane component $\sigma_z$ of the Pauli matrices vector in the outer Fermi arc for each strain. $\theta$ is measured anticlockwise and defined as an angle in the polar coordinate system where its reference point is $\Gamma$-point and reference direction is the [100]-direction ($\Gamma\to$X).}\tabularnewline
\vspace{-6ex}
\end{figure}

In order to obtain the information of the spin-orbit Hamiltonian
inducing the spin splitting, we calculate spin textures shown in Fig. 3.
For the unstrained case (0\%), our calculated spin texture shown in Fig.
3(a) shows a typical Rashba spin splitting, which is an agreement with
the previous study \cite{Nishida_First_2014}. As the tensile strain
increases, however, the nature of the spin texture changes dramatically.
For the tensile strain (See Fig. 3(b), (c)), the spin texture is
different from the Rashba-type one, and also has the nonzero
out-of-plane component that can be considered to originate from the
spin-orbit Hamiltonian
\(H_{[1\bar10]z}^\perp=\alpha_{[1\bar10]z}^\perp\left(\frac{k_x-k_y}{\sqrt2}\right)\sigma_z\).
In this case, the out-of-plane (\(\sigma_z\)) component is almost
constant for the k-points except part of the Fermi arc around the
degenerate points in the {[}110{]}-direction
(\(\theta=45^\circ, -135^\circ\)) so that the electronic states may have
a long spin lifetime. This suggests that the persistent spin helix (PSH)
\cite{Bernevig_Exact_2006, Koralek_Emergence_2009, Absor_Persistent_2015}
may be formed. PSH has one-dimensionally-oriented spin components and is
important for spin field-effect transistor (spinFET) \cite{Datta_Electronic_1990} due to its long spin
lifetime \cite{Bernevig_Exact_2006}. As shown in Fig. 2(c), the spin
components are one-dimensionally oriented at the tensile strain of 5\%.
The estimated PSH period \(\lambda_{PSH}\) is
\(\pi/k_{[1\bar10]}=0.098\ \mu\)m, which is comparable with that for a
ZnO \((10\bar10)\) surface (0.19 \(\mu\)m) \cite{Absor_Persistent_2015},
and is two order of magnitude smaller than that for a GaAs/AlGaAs
quantum well (7.3-10 \(\mu\)m) \cite{Walser_Direct_2012}. Therefore, the
PSH state of strained LaAlO\textsubscript{3}/SrTiO\textsubscript{3} is
suitable for miniaturization of spintronic devices.

In summary, we have performed first-principles density functional
calculations for the n-type interface in LaAlO\(_3\)/SrTiO\(_3\) with
the tensile strain. We found that the spin-orbit coupling coefficient
\(\alpha\) directly connected to IREE could be controlled by the tensile
strain and enhanced up to 5 times for the tensile strain of 7\%. The PSH
may be induced for the tensile strain so that the extremely long spin
lifetime can be achieved. Moreover, the estimated PSH period
\(\lambda_{PSH}\) is comparable with that for a ZnO \((10\bar10)\)
surface and is also two order of magnitude smaller than that for a
GaAs/AlGaAs quantum well \cite{Walser_Direct_2012}. The 2DEG in oxide
surfaces \cite{Santander-Syro_Two_2011, King_Subband_2012} and
interfaces \cite{Yang_High_2016} may be a promising candidate for
miniaturization of spintronic devices utilizing PSH. These results
support that the strain effect in LaAlO\(_3\)/SrTiO\(_3\) is important
for various applications such as spinFET and spin-to-charge conversion.

\acknowledgment
The authors thank H. Kotaka for invaluable discussion about analyzing
spin textures. This work was supported by JSPS KAKENHI Grant Numbers
JP25790007, JP15H01015, JP17H05180. This work was also supported by
Kanazawa University SAKIGAKE Project. The computation was mainly carried
out using the computer facilities at RIIT, Kyushu University. This work
was supported in part by MEXT as a social and scientific priority issue
(Creation of new functional devices and high-performance materials to
support next-generation industries) to be tackled by using post-K
computer (Project ID: hp160227).
\bibliographystyle{jjap}
\bibliography{readcube_export}
\end{document}